\input{aipcheck}

\documentclass[
    ,final            
  ]
  {aipproc}

\layoutstyle{6x9}

\newcommand{\oversim}[2]{\protect{\mbox{\lower0.5ex\vbox{%
   \baselineskip=0pt\lineskip=0.2ex
   \ialign{$\mathsurround=0pt #1\hfil##\hfil$\crcr#2\crcr\sim\crcr}}}}} 
\newcommand{\simgreat}{\mbox{$\,\mathrel{\mathpalette\oversim>}\,$}} 

\begin{document}

\title{Are stellar over-densities in dwarf galaxies the ``smoking gun'' of triaxial dark matter haloes?}

\classification{<\texttt{http://www.aip..org/pacs/index.html}>}

\author{Jorge Pe\~narrubia}{
  address={Institute of Astronomy, University of Cambridge, Madingley Road, Cambridge, CB3 0HA, UK}
}
\author{Matthew G. Walker}{
  address={Institute of Astronomy, University of Cambridge, Madingley Road, Cambridge, CB3 0HA, UK} 
}
\author{Gerard Gilmore}{
  address={Institute of Astronomy, University of Cambridge, Madingley Road, Cambridge, CB3 0HA, UK}
}
\begin{abstract}
 We use N-body simulations to study the tidal evolution of globular clusters (GCs) in dwarf spheroidal (dSph) galaxies. Our models adopt a cosmologically motivated scenario in which the dSph is approximated by a static NFW halo with a triaxial shape.  For a large set of orbits and projection angles we examine the spatial and velocity distribution of stellar debris deposited during the complete disruption of stellar clusters.  Our simulations show that such debris appears as shells, isolated clumps and elongated over-densities at low surface brightness ($\geq 26$ mag/arcsec$^2$), reminiscent of substructure observed in several MW dSphs. Such features arise from the triaxiality of the galaxy potential and do {\it not} dissolve in time. Stellar over-densities reported in several MW dSphs may thus be the telltale evidence of dark matter haloes being triaxial in shape. 
 We explore a number of kinematical signatures that would help to validate (or falsify) this scenario. The mean angular momentum of the cluster debris associated with box and resonant orbits, which are absent in spherical potentials, is null. As a result, we show that the line-of-sight velocity distribution may exhibit a characteristic ``double-peak'' depending on the oriention of the viewing angle with respect to the progenitor's orbital plane. Kinematic surveys of dSphs may help to detect and identify substructures associated with the disruption of stellar clusters, as well as to address the shape of the dark matter haloes in which dSphs are embedded. 

\end{abstract}

\maketitle


\section{Introduction}
Dwarf spheroidal (dSph) galaxies are the most dark-matter (DM) dominated galaxies known to date (see reviews by Mateo 1998 and Gilmore et al. 2007).  Dynamical constraints on the DM distribution in dSphs may therefore be compared directly with the predictions from cosmological simulations to study the elusive nature of dark matter.

In the last few years, however, wide-field imaging and multi-object spectroscopic surveys have revealed that dSph stellar components are more complex than previously believed. Localized oddities such as shells (Coleman et al. 2005; Olszewski et al. 2006), butterfly-shaped substructures (Stetson \& Hesser 1998) and cold clumps (Kleyna 2003; Ibata et al. 2006)
have been reported in several dSphs. These results conflict with the naive expectation that dSph DM haloes (with typical crossing times of a few hundred Myr) should host well-mixed stellar populations, and may thus pose a challenge to Cold Dark Matter (CDM).

A second oddity in this context is the existence of globular clusters (GCs) in dSphs. Although GCs are considerably more common in dwarf irregular galaxies (e.g. Georgiev et al. 2008), at least two of the Milky Way dSphs---Fornax and Sagittarius---contain GCs (five and four, respectively). The formation of GCs in a cosmological context still lacks theoretical understanding, and it is unclear whether GCs could also have formed in other dSphs, but not survived to the present epoch. The present contribution elaborates on this scenario by studying the possible imprints that tidally disrupted clusters may leave in the stellar populations of dSphs. In this contribution we provide a summary of the results found by Pe\~narrubia, Walker \& Gilmore (2009). We refer interested readers to that paper for details.

\section{Models and Numerical Setup}

Our models assume that the mass profile of a dSph can be approximated by a {\it triaxial} Navarro, Frenk \& White (1997, NFW for short) DM halo. We choose axis-ratios $c/a=0.5$; $b/a=0.83$, typically found in cosmological simulations (e.g. Jing \& Suto 2002). We scale the free parameters of the NFW potential to those typically estimated in Milky Way dSphs, a peak velocity $v_{\rm max}=20.6$ km/s and a peak radius $r_{\rm max}=4.1$ kpc (Pe\~narrubia et al. 2008a,b).

Our GC models are N-body realizations of a King (1966) profile orbiting within a static NFW potential. For illustrative purposes, we choose a GC model that resembles the cluster F1 of the Fornax dSph, which has a mass of $3.7\times 10^4 M_\odot$, and core and tidal radii of $R_c=10$ pc and $R_t=60$ pc, respectively.
 Dynamical friction between the GC and the background DM particles is introduced in our computations as an external force term following the method outlined in Pe\~narrubia et al. (2006). 

We follow the evolution of the cluster N-body model in the dwarf potential using {\sc Superbox}, a highly efficient particle-mesh gravity code. Our grid resolution is $0.16 R_c$ and the time step adopted is $dt=t_{\rm cr}/25$, where $t_{\rm cr}=15.7$ Myr is the crossing time of our cluster model.

\section{Results}
Triaxial potentials allow an infinite number of orbits, which are typically classified in four major families  (i) boxes, (ii) loops (iii) resonant (iv) irregular orbits. Their phase-space structure show remarkable differences (e.g. Binney \& Tremaine 2008). 

In this contribution we only consider clusters orbiting in the $X-Z$ symmetry planes; recall that the coordinates $(X,Y,Z)$ are oriented along the major, intermediate and minor axis of the potential, respectively. We emphasize that our orbital sample is a largely reduced subset of the whole orbital family spectrum allowed by the potential adopted in our study. 

\begin{figure*}
\includegraphics[height=.45\textheight]{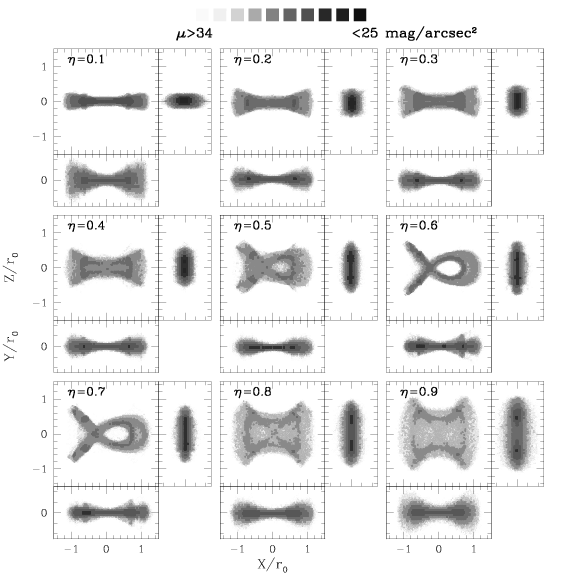}
\caption{Projected surface brightness associated with the tidal debris from the cluster F1 on different orbits initially confined on the symmetry plane formed by the major and minor axis of the potential. The initial apocentric distance is $r_0=0.5$ kpc and $\eta$ is the orbital circularity. For each model we show the projection along the three principal axis. } 
\label{fig:SBXZ}
\end{figure*}

\subsection{Debris morphologies in triaxial potentials}
In Fig.~\ref{fig:SBXZ} we show a few examples of tidal debris configurations that one would expect from a fully disrupted F1 cluster moving on orbits with different circularity ($\eta$).

The projected morphologies are extremely rich. 
By colour-coding cluster particles according to their projected surface brightness we can observe the formation of ``folds'' and ``cusps`` in regions densely populated by debris material. These over-densities may produce striking features. For example, folds associated with box orbits exhibit a ``butterfly''-like shape (e.g. $\eta=0.8$). Even more complex do appear the stellar substructures that arise from the disruption of clusters on resonant orbits. A good example is the ``fish'' orbit in this Figure ($\eta=0.5$). Note, however, the strong dependence of debris morphology on the line-of-sight projection.

Our models show that stellar substructures arise naturally from the disruption of GCs in a static potential. 
Shells (folds) and isolated over-densities (cusps) arise in almost all orbital configurations and line-of-sight projections. These features result from the triaxial shape of the galaxy potential and do {\it not} dissolve in time. 
\begin{figure}
\includegraphics[height=.32\textheight]{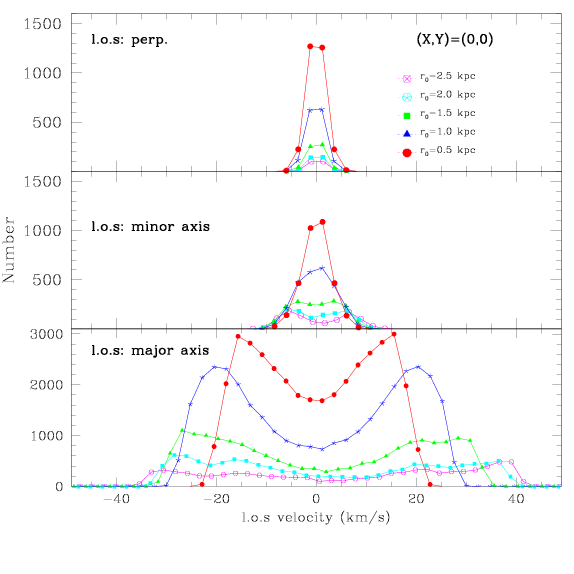}
\caption{Projected line-of-sight velocity distributions of orbits with different apocentres ($r_0$) measured at the centre of the dwarf. The initial orbital circularity in these models is $\eta=0.8$.} 
\label{fig:vdist}
\end{figure}

\subsection{Kinematic Signatures}
The tidal disruption of a $10^4 L_\odot$ globular cluster leaves stellar substructures with surface brightnesses $\mu\simgreat 26$ mag/arcsec$^2$. Some of these features fall within the detection limits of present instrumentation. However, in areas where dwarf field stars dominate in number, the identification of cluster debris as stellar over-densities may prove challenging. Kinematic surveys may provide additional constraints.

Box, resonant and irregular orbits, {\it all absent in spherical potentials}, have null angular momentum. In practice this implies that, at any position angle, there is the same amount of debris material moving away from us, as towards us. This results in a line-of-sight velocity distribution that is always symmetric and centred at $\langle v_{l.o.s}\rangle =0$, but which may show distinctive ``double-peaks'' depending on the oriention of the viewing angle with respect to the progenitor's orbital plane, as shown in Fig.~\ref{fig:vdist}. Such velocity distributions would not be expected in well-mixed stellar populations.
Kinematic surveys of dSphs are thus useful tools to identify substructures associated to the disruption of stellar clusters, as well as to address the shape of DM haloes.



\bibliographystyle{aipprocl} 

%


\end{document}